\documentclass[aoas,preprint]{imsart}

\RequirePackage[OT1]{fontenc}
\RequirePackage{amsthm,amsmath}
\RequirePackage{natbib}
\usepackage{graphicx,psfrag,epsf}
\usepackage{threeparttable}
\usepackage{bm}
\usepackage[toc,page]{appendix}

\startlocaldefs
\numberwithin{equation}{section}
\theoremstyle{plain}

\endlocaldefs

\begin{document}

\begin{frontmatter}
\title{Effective Model Calibration via Sensible Variable Identification and Adjustment, with Application to Composite Fuselage Simulation}
\runtitle{Calibration of Sensible Variables}
%

\begin{aug}
\author{\fnms{Yan } \snm{Wang}\thanksref{m1}\ead[label=e1]{yanwang@bjut.edu.cn}},
\author{\fnms{Xiaowei } \snm{Yue}\thanksref{m2}\ead[label=e2]{xwy@vt.edu}},
\author{\fnms{Rui } \snm{Tuo}\thanksref{m3}\ead[label=e3]{ruituo@tamu.edu}},
\author{\fnms{Jeffrey } \snm{H. Hunt}\thanksref{m4}\ead[label=e4]{jeffrey.h.hunt@boeing.com}},
\and
\author{\fnms{Jianjun} \snm{Shi}\thanksref{m5}
\ead[label=e5]{jianjun.shi@isye.gatech.edu}}
\affiliation{Beijing University of Technology\thanksmark{m1}, Virginia Polytechnic Institute and State University \thanksmark{m2}, Texas A \&M University\thanksmark{m3}, The Boeing company \thanksmark{m4} and Georgia Institute of Technology \thanksmark{m5}}

\address{Yan Wang \\
College of Applied Sciences \\
Beijing University of Technology\\
Beijing 100124\\
China \\
\printead{e1}\\
\phantom{E-mail:\ }}

\address{Xiaowei Yue \\
Grado Department of Industrial \& Systems Engineering\\
Virginia Polytechnic Institute and State University\\
VA 24061\\
USA\\
\printead{e2}\\
\phantom{E-mail:\ }}

\address{Rui Tuo \\
Department of Industrial \& Systems Engineering\\
Texas A\&M University College Station\\
TX 77843\\
USA\\
\printead{e3}\\
\phantom{E-mail:\ }}

\address{Jeffrey H. Hunt\\
The Boeing company\\
CA 90245\\
 USA\\
\printead{e4}\\
\phantom{E-mail:\ }}

\address{Jianjun Shi\\
The H. Milton Stewart School of Industrial and Systems Engineering\\
 Georgia Institute of Technology\\
 GA 30332\\
 USA\\
\printead{e5}\\
\phantom{E-mail:\ }}

\end{aug}

\begin{abstract}
Estimation of model parameters of computer simulators, also known as calibration, is an important topic in many engineering applications. In this paper, we consider the calibration of  computer model parameters with the help of engineering design knowledge. We introduce the concept of sensible (calibration) variables. Sensible variables are model parameters which are sensitive in the engineering modeling, and whose optimal values differ from the engineering design values.
We propose an effective calibration method to identify and adjust the sensible variables with limited physical experimental data. 
 The methodology is applied to a composite fuselage simulation problem.
\end{abstract}



\end{frontmatter}

\section{Introduction}
\label{sec:intro}
Composite parts have been widely used in various industrial applications including aerospace, automotive, energy industries due to their superior properties including high strength-to-weight ratio, high stiffness-to-weight ratio, potentially long life usage and low life-cycle cost \citep{mallick2007fiber}. 
  Dimensional variation modeling of composite parts lays a foundation for quality control and process improvement, and numerous studies have been conducted in this area. Literature review related to composite parts assembly and variation modeling refers to 
   \cite{shi2006stream}, \cite{zhang2016stream,zhang2016stream2} and \cite{yue2017surrogate}. Most of these studies in variation modeling and analysis of composite parts are based on the finite element analysis (FEA). FEA is a numerical simulation method and widely used in structural analysis, fluid dynamics, heat transfer, and electromagnetic potential analysis. Therefore, obtaining an accurate finite element model is a fundamental step for subsequent variation analysis and quality control of composite fuselage assembly.

 Finite element modeling of composite fuselage is a challenging task due to the compliant nature and anisotropic characteristics of composite structures \citep{jones1998mechanics,  barbero2013finite}. Many model parameters, such as material property parameters, multiple layer thickness, and constraints on composite parts, make important impacts on the accuracy of the computer model. Although the values of these parameters are available according to engineering design knowledge, the actual values of those parameters in the real fabricated fuselage are unknown because of inevitable variabilities in the manufacturing system, such as part fabrication errors, fixture errors, and positioning errors.

The present research is motivated by a composite fuselage dimensional simulation.
Our goal is to find the optimal values of the model parameters, under which the finite element outputs match the structural load experimental observations of the composite fuselage. In the area of computer experiments, identification of unknown model parameters based on computer simulation outputs and physical experimental observations is referred to as calibration of computer models. There has been a vast amount of literature discussing the calibration methodologies and their applications since the pioneer work by \cite{kennedy2001bayesian} appeared. To name a few authors, see \cite{kennedy2001bayesian}, \cite{higdon2004combining}, \cite{bayarri2012framework}, \cite{joseph2015engineering}, \cite{tuo2015efficient}, and \cite{gramacy2015calibrating}.
Calibration of computer models has been applied successfully in many areas, like the research of hydrocarbon reservoir \citep{craig2001bayesian}, the electrical activity of myocytes in cell biology \citep{plumlee2015calibrating},  the diffusion of radionuclides released during nuclear bomb tests \citep{pratola2016bayesian} and so on.

In a typical computer experiment problem, the corresponding physical experiment is expensive to run, and this is why we need the help of the computer simulator. Thus, the physical experimental sample size is normally rather limited. On the other hand, many computer simulators have a number of calibration parameters. In many problems, simultaneously estimating all calibration parameters from the data can be intractable due to the \textit{curse of dimensionality}.
In the composite fuselage simulation problem discussed in Section \ref{sec:case}, there are quite a few calibration parameters, such as material property parameters, multiple layer thickness, support parameters, fabrication angle, and temperature. However, only eight physical experimental observations are obtained because of the time and cost constraints. 

To break the curse of dimensionality, we resort to the engineering knowledge to obtain a sparse model for the calibration problem.
In most calibration problems in engineering, engineers have some expert knowledge of the model parameters using the information from the design process. Based on the knowledge, engineers can give some engineering values for the calibration parameter. Because the engineering knowledge is likely to be reliable, most engineering values are close to their corresponding optimal values and only a subset of the model parameters need to be adjusted. We call them the \emph{sensible calibration variables}, or abbreviated as \emph{sensible variables}. Adjusting the sensible variables can significantly improve the accuracy of the simulation model. The sensible variables differ from the sensitive variables which have been widely considered in engineering problems. The related area is known as sensitivity analysis \citep{shi2006stream}. 

In this paper, we propose an effective method to identify and adjust the sensible variables. 
A loss function of the projected kernel calibration method introduced by \cite{tuo2017projected} is used to measure the model fitness, {{and an $l_1$-type penalty \citep{tibshirani1996regression} is used to encourage a minimum adjustment for the calibration parameters}}. We employ an approximation technique and the objective function becomes convex and thus can be solved efficiently.
The proposed method proceeds by identifying and adjusting the sensible variables, which is particularly useful when the physical experimental observations are limited.
The performance of the proposed method is shown in numerical studies. We apply the proposed method to the calibration of the finite element model in the composite fuselage simulation problem.


 The remainder of this paper is organized as follows. In Section \ref{sec:rev}, we review some existing calibration methods related to the current work. In Section \ref{sec:meth}, we propose an effective model calibration via sensible variable identification and adjustment. 
 In this section, we introduce the concept of sensible variables and discuss their properties. We also illustrate how the proposed method can identify and adjust the sensible variables.
  In Section \ref{sec:case}, we study the finite element model calibration problem of composite fuselage using the proposed method. Concluding remarks and further discussion are given in Section \ref{sec:ext}. The data from the composite fuselage simulation problem and the R code are provided in the supplementary material.

\section{Review on Calibration of Computer Models}
\label{sec:rev}
Calibration of computer models has attracted considerable attention in the literature.
In a calibration problem, the input of the computer model consists of two types of variables: the control variables $\mathbf x=(x_1,\ldots,x_d)^{T}\in \mathbf R^d$ and the calibration parameters $\bm\theta=(\theta_1,\ldots,\theta_m)^T\in\mathbf R^m$. The control variables are variables that can be controlled in the corresponding physical experiments. The calibration parameters are variables that are involved in the computer models but their values cannot be controlled or measured in the physical experiments. Normally, calibration parameters represent certain inherent attributes of the physical system. We refer to \cite{kennedy2001bayesian} for more discussion about the calibration parameters.
Denote the set of design points for the physical experiments by $\{\mathbf{x}_1,\ldots,\mathbf x_{n}\}$, and the corresponding responses by $\mathbf{Y}=\{y_1,\ldots,y_n\}$.
A commonly used model for the physical experimental observations is
\begin{eqnarray}
y_i=\zeta(\mathbf x_i)+e_i,
\label{eq2.1}
\end{eqnarray}
where $i=1,\ldots,n$. $\zeta(\cdot)$ is called the \emph{true process}, which is an unknown function, and $e_i$'s are the observation errors following $N(0,\sigma^2)$ with unknown $\sigma^2<\infty$.
{{\cite{kennedy2001bayesian} claim that the computer outputs cannot perfectly fit the physical experimental observations because the computer outputs are biased: the computer models are usually built under assumptions and simplifications which are not exactly correct in reality.
Taking this model bias into account, one could use the following model \citep{kennedy2001bayesian,higdon2004combining}
\begin{eqnarray}
\zeta(\cdot)= y^s(\cdot,\bm\theta^*)+\delta(\cdot),
\label{eq2.2}
\end{eqnarray}
where $\bm\theta^*$ denotes the combination of the optimal calibration parameters, {{$y^s$ denotes the computer model}} and $\delta$ is the \textit{discrepancy function}. The goal of calibration is to estimate $\bm\theta^*$, so that the computer outputs are close to the physical experimental observations. However, equation (\ref{eq2.2}) is not enough to fully determine $\bm\theta^*$, because the function $\delta$ is also unknown. This problem is known as the identifiability issue of Kennedy and O'Hagan's method.}}

{{To resolve this identifiability problem,  \cite{tuo2015efficient} define $\bm\theta^*$ as}}
\begin{eqnarray}
\begin{aligned}
\bm\theta^{*}:=\operatorname*{argmin}_{\bm\theta}\int_{\Omega}   ({\zeta(\mathbf x)-y^s(\mathbf x,\bm\theta)})^2 d\mathbf{x},
\end{aligned}
\label{thetastar}
\end{eqnarray}
{{i.e., $\bm\theta^{*}$ minimizes the $L_2$ distance
between the true process and the computer outputs.
\cite{plumlee2016bayesian} observes that (\ref{thetastar}) is generally equivalent to 
the orthogonality constraints}}
 \begin{eqnarray}
\int_{\Omega}\frac{\partial{y^s(\mathbf x,\bm\theta^{*})}}{\partial{\theta_j}}\delta(\mathbf x)d\mathbf x=0,
\label{eq2.8}
\end{eqnarray}
for $j=1,\ldots,m$. Based on (\ref{eq2.8}), \cite{plumlee2016bayesian} proposes a Bayesian calibration method which rectifies the identifiability problem of Kennedy and O'Hagan's model.

Inspired by the work of \cite{plumlee2016bayesian}, \cite{tuo2017projected} suggests a frequentist method to estimate the calibration parameter. {{First, we choose a positive definite kernel function $\Phi$}}. A common choice is the Gaussian correlation family with
\begin{eqnarray}
\Phi(\mathbf{x}_i,\mathbf{x}_j)=\exp(-\phi {\parallel \mathbf{x}_i-\mathbf{x}_j\parallel}^2),
\label{eq2.6}
\end{eqnarray}
for some $\phi>0$. The projected kernel of $\Phi$ according to the constraints (\ref{eq2.8}) is defined as
\begin{eqnarray}
 \Phi_{\bm\theta}(\mathbf x_i,\mathbf x_j)=\Phi(\mathbf x_i,\mathbf x_j)-h_{\bm\theta}(\mathbf x_i)^{T}H_{\bm\theta}^{-1}h_{\bm\theta}(\mathbf x_j),
\label{eq2.10}
\end{eqnarray}
with
 \begin{eqnarray}
 \begin{gathered}
h_{\bm\theta}(\mathbf x)=\int_{\Omega}\frac{\partial{y^s(\mathbf x',\bm\theta)}}{\partial{\bm\theta}}\Phi(\mathbf x',\mathbf x)d\mathbf x',\\
H_{\bm\theta}=\int_{\Omega}\int_{\Omega}\frac{\partial{y^s(\mathbf x',\bm\theta)}}{\partial{\bm\theta}}\left\{\frac{\partial{y^s(\mathbf x',\bm\theta)}}{\partial{\bm\theta}}\right\}^{T}\Phi(\mathbf x',\mathbf x)d\mathbf x'd\mathbf x.
\end{gathered}
\label{eq2.11}
\end{eqnarray}
 \cite{tuo2017projected} proposes the following projected kernel calibration estimator
  \begin{eqnarray}
\hat{\bm\theta}^{PK}=\operatorname*{argmin}_{\bm\theta}(\mathbf Y-\mathbf{Y}^s_{\bm\theta})^{T}(\mathbf \Phi_{\bm\theta} +\eta^2 I_n)^{-1}(\mathbf Y-\mathbf{Y}^s_{\bm\theta}),
\label{eq2.12}
\end{eqnarray}
where $\mathbf{Y}^s_{\bm\theta}=\{y^s(\mathbf{x}_1,\bm\theta),\ldots,y^s(\mathbf{x}_{n},\bm\theta)\}^{T}$,
  $ \mathbf \Phi_{\theta}=[\Phi_{\theta}(\mathbf x_i,\mathbf x_j)]_{1\leq i,j\leq n}$, $I_n$ denotes the identity matrix and $\eta>0$ is a tuning parameter which can be chosen using the Bayesian method suggested by \cite{chang2014model}. 
%
{ The scale parameter $\phi$ can be estimated by the maximum likelihood (ML) method, i.e., in (\ref{eq2.12}) we maximize the objective function with respect to both $\bm \theta$ and $\phi$.}
\cite{tuo2017projected} proves the consistency and efficiency of the projected kernel estimator.



Because of the high cost of physical experimental runs, only a small number of physical experimental observations can be obtained. Thus, only a few model parameters can be adjusted using a data-driven method. However, in many engineering problems, the number of model parameters can be large so that not all of them can be adjusted effectively.
In the composite fuselage FEA simulation problem in Section \ref{sec:case}, the computer model has one control variable (actuator force) and five calibration parameters, including surface body thickness, support parameter, material thickness ratio, fabrics orientation, and temperature. However, only eight physical experiments can be conducted by adjusting actuator forces to collect physical experimental observations. Similar challenges of parameter estimation occur in many composite parts finite element model calibration problems in the areas of aerospace, automotive, energy industries, etc. Clearly, existing calibration methods will suffer from the curse of dimensionality.

To break the curse of dimensionality, we will introduce the concept of sensible variables in Section \ref{sec:meth}. An effective calibration method will be proposed to solve the calibration problems with limited physical experimental observations.



\section{Methodology}
\label{sec:meth}

In this section, we introduce a novel methodology which tackles the curse of dimensionality in calibration problems with the help of engineering design information. A {variable selection and estimation procedure} is performed by identifying a subset of the calibration parameters which need to be adjusted. Such calibration parameters are called \textit{sensible variables}. {{ The most common definitions of “sensible” are: practical, reasonable, logical, rational, etc. Such as \textit{sensible heat} is literally the heat that can be felt. It is the energy from one system to another that change the temperature. Sensible heat is used in contrast to latent heat, which is the heat changing the phase. 
In the proposed method,  by identifying and adjusting these sensible parameters, the performance of the computer model will be improved, that is the discrepancy between the computer model and the physical experimental observations will be reduced. }}

\subsection{Sensible Variable}
\label{sec:sensible}

{ In a typical engineering problem, initial guesses of the model parameters are usually available. These values are generally obtained using the engineering design information or the domain expert knowledge. We call them the \textit{engineering design values}.}
Ideally, the physical properties of a product should be consistent with its engineering design values. This consistency, however, may be violated in practice due to certain inevitable variability in the manufacturing system, such as fabrication errors, fixture errors. In the composite fuselage simulation, although an ideal engineering design value of the surface body thickness is given, the actual thickness of a specific batch of the fabricated fuselage is still unknown because of fabrication errors. From engineering design, the fabrics orientation should be 45 degrees but fabrication uncertainty may also exist. 


 {
 Recall that calibration of model parameters may suffer from the curse of dimensionality when the input dimension of the calibration parameters is relatively high.} 
 Fortunately, our engineering knowledge suggests that we can reasonably assume that most of the calibration parameters can be set as their engineering design values, because the quality of the product is generally well controlled. Thus, only a small number of the model parameters need to be adjusted.
 We call these variables as the \emph{sensible calibration variables}, or abbreviated as \textit{sensible variables}.
For the remaining calibration parameters, which do not need to be adjusted, we call them \textit{insensible variables}.

It is worth noting that sensible variables differ from sensitive variables. The latter has been widely used in engineering modeling, which are model variables that have significant influence on the simulation outputs.  
{{Sensible variables must be sensitive. If a model parameter is insensitive, the right hand side of (\ref{thetastar}) is largely unaffected by this parameter, and the optimization with respect to this paramter in (\ref{thetastar}) does not make sense. On the other hand, sensitive variables are not necessarily sensible. When the engineering design value of a model parameter is close to its optimal value, we do not need to adjust this parameter, even if it is sensitive.

Denote the engineering design values as  $\bm\theta^{(0)}=(\theta_1^{(0)},\ldots,\theta_m^{(0)})^{T}$.  We summarize the proceeding discussion on the relationship between sensibility and sensitivity in Table \ref{Table 1}:
%
\begin{table}[htpb]
	\begin{center}
		\caption{Sensibility versus sensitivity.}
		\label{Table 1}
		\begin{tabular}{c|c|c}
			\hline
			Variable Type & Description &Needs Adjustment?\\
			\hline
			Insensitive            &  {$\theta_i$ has less influence on the simulation outputs}   &   No\\
			 Sensitive but insensible & {$\theta_{i}$ is sensitive and $\theta_i^{(0)}=\theta_i^*$ }&   No \\
			Sensible  & {$\theta_{i}$ is  sensitive and $\theta_i^{(0)}\neq \theta_i^*$  }           &   Yes\\
			\hline
		\end{tabular}
	\end{center}
\end{table}

  }}


In the next two subsections, we will propose a statistical method to identify and adjust the sensible variables.


\subsection{Surrogate Modeling of Computer Models}\label{sec:surrogate}

Before introducing the proposed calibration methodology, we consider the surrogate modeling of the computer outputs with many calibration parameters. In the fuselage simulation, each run of the FEA code is time consuming. Thus it is unrealistic to run the code as many times as we want. 
In order to make statistical inference about the model parameters, we need to reconstruct the computer output response surface based on computer code runs over a set of designed input settings. Specifically, we choose a set of design points $\{(\mathbf x_1, \bm\theta_1),\ldots, (\mathbf x_N, \bm\theta_N)\}$ and run the computer code over each point in this set. Space-filling designs are commonly used for this purpose in the computer experiments literature. We refer to \cite{santner2013design} for a review of computer experiment design methods. Based on $(\mathbf x_i,\mathbf{\theta}_i,y^s(\mathbf x_i,\mathbf{\theta}_i))_{i=1}^N$, we build a surrogate model $\hat{y}^s$, which serves as an approximation to $y^s$.

Recall that $\mathbf x=(x_1,\ldots,x_d)$ and $\bm\theta=(\theta_1,\ldots,\theta_m)$. 
Assume that $y^s$ is nearly linear in $\bm\theta$.  We apply a Taylor expansion to $y^s(\mathbf x,\bm\theta)$ at $\bm\theta^{(0)}$ and obtain
\begin{multline}
\label{eq3.3}
y^s(\mathbf x,\bm\theta)=y^s(\mathbf x,\bm\theta^{(0)})+\sum_{i=1}^m\frac{\partial y^s(\mathbf x,\bm\theta^{(0)})}{\partial {\theta_i}}(\theta_i-\theta_i^{(0)})
+O(\|\bm\theta-\bm\theta^{(0)}\|_2^2).
\end{multline}
We assume that the residual term is negligible.
Then we can consider the surrogate model $\hat y^s(\mathbf x,\bm\theta)$ with the form
\begin{eqnarray}
\hat y^s(\mathbf x,\bm\theta)=\hat{f}(\mathbf x)+\bm\theta^{T}\hat{\mathbf g}(\mathbf x),
\label{eq3.4}
\end{eqnarray}
where $\hat{f}(\mathbf x)$ and $\hat{\mathbf g}(\mathbf x)$ can be regarded as estimates of $y^s(\mathbf x,\bm\theta^{(0)})-\frac{\partial y^s(\mathbf x,\bm\theta^{(0)})}{\partial {\bm\theta}}\bm\theta^{(0)}$ and $\frac{\partial y^s(\mathbf x,\bm\theta^{(0)})}{\partial {\bm\theta}}$ respectively. 
It is worth noting that $\hat{f}$ and $\hat{\mathbf g}$ are independent of the parameter $\bm\theta$.{ {This implies that the complex computer models have been approximated by models in linear form.}}

Another advantage of surrogate model (\ref{eq3.4}) is that $\frac{\partial \hat{y}^s}{\partial \bm\theta}=\hat{\mathbf g}$ is independent of $\bm\theta$. As a result, the matrix $\Phi_{\bm\theta}$ defined in (\ref{eq2.10}) is also independent of $\bm\theta$ and thus (\ref{eq2.12}) becomes a convex optimization problem. In Section \ref{sec:calibration}, we will use (\ref{eq3.4}) and formulate the penalized calibration as a convex optimization problem which can be solved efficiently.

Now we consider how to construct the surrogate model (\ref{eq3.4}) given the data from the computer experiment. A simple method is to impose parametric models on $\hat{f}$ and $\hat{\mathbf g}$. For instance, we may suppose that $\hat{f}$ and $\hat{\mathbf g}$ are linear in $x$, and then the parameters can be estimated using the least squares method. We will adopt this parametric method in the case study discussed in Section \ref{sec:case}. 

{{A more flexible method can be consider, by using Gaussian process modeling to construct $\hat{f}$ and $\hat{\mathbf g}$. 
Write $y^s_i=y^s(\mathbf x_i,\mathbf{\theta}_i)$. Then we consider the following Gaussian process surrogate model
\begin{eqnarray}\label{GPmeta}
y^s_i=f(\mathbf x_i)+\bm\theta_i^T \mathbf g(\mathbf x_i)+\varepsilon_i,
\end{eqnarray}
where $\mathbf g=(g_1,\ldots,g_m)^T$; $f,g_1,\ldots,g_m$ are modeled as realizations of independent Gaussian processes $F, G_1,\ldots,G_m$, respectively; and $\varepsilon_i$ are mutually independent nugget variables following the normal distribution $N(0,\tau^2)$ with unknown $\tau^2>0$. The simplest Gaussian process models for $f$ and $\mathbf g$ are the simple kriging models, in which we assume $F, G_1,\ldots, G_m$ have zero mean and covariance functions $\kappa^2 K(\cdot,\cdot),\kappa^2_1 K_1(\cdot,\cdot),\ldots,\kappa^2_m K_m(\cdot,\cdot)$ respectively, with unknown variances $\kappa^2,\kappa^2_1,\ldots,\kappa_m^2>0$ and known correlation functions $K,K_1,\ldots,K_m$. One may introduce more degrees of freedom to these simple kriging models so that more complex functions can be fit. Such extensions are fairly straightforward and have been widely discussed in the literature. See \cite{santner2013design,banerjee2004hierarchical} for example. The nugget term $\varepsilon_i$ is necessary. This is because the partial linear Gaussian process model $f(\mathbf x)+\mathbf{\theta}^T \mathbf g(\mathbf x)$ is only an approximate to the true function $y^s(\mathbf x,\bm\theta)$ and thus $f(\mathbf x_i)+\mathbf{\theta}_i^T \mathbf g(\mathbf x_i)$ may not be able to interpolate $y_i^s$ for all $i$.

The functions $f$ and $\mathbf g$ can be estimated using the standard conditional inference technique for Gaussian process surrogate models. Conditional on the observed data and the model parameters, the expectations of $F(\mathbf x)$ and $\mathbf G(\mathbf x)=(G_1(\mathbf x),\ldots, G_m(\mathbf x))$ are
\begin{eqnarray}\label{f}
a^T (C+\tau^2 I_N)^{-1}\mathbf Y_N^s,
\end{eqnarray}
and
\begin{eqnarray}\label{g}
B(C+\tau^2 I_N)^{-1}\mathbf Y_N^s.
\end{eqnarray}
respectively, where $\mathbf Y_N^s=(y^s_1,\ldots,y^s_N)^T$, $a=(\kappa^2 K(\mathbf x_1,\mathbf x),\ldots,\kappa^2 K(\mathbf x_n,\mathbf x))^T$, $B=[\kappa_j^2 K_j(\mathbf x_i,\mathbf x)]_{i j}$, $C=[\kappa^2 K(\mathbf x_i,\mathbf x_j)+\sum_{l=1}^m\theta_{li}\theta_{lj} \kappa_l^2 K_l(\mathbf x_i,\mathbf x_j)]_{i j}$ and $I_N$ denotes the identity matrix. We can choose $\hat{f}$ and $\hat{\mathbf g}$ in (\ref{eq3.4})
as (\ref{f}) and (\ref{g}) respectively. The model parameters can be estimated using the maximum likelihood method or the Bayesian methods. For the parameter estimation for Gaussian process models, we refer to \cite{santner2013design,banerjee2004hierarchical}. A related Gaussian process surrogate model is considered by \cite{ba2012composite}.}}

\subsection{Penalized Orthogonal Calibration}
\label{sec:calibration}

Recall that the calibration parameter is $\bm\theta=(\theta_1,\ldots,\theta_m)^T$. As discussed in Section \ref{sec:sensible}, the key to break the curse of dimensionality is to adjust only the sensible variables, which is an \textit{unknown} subset of $\{\theta_1,\ldots,\theta_m\}$. As before, we denote the engineering design parameter values by the vector $\bm\theta^{(0)}=(\theta_1^{(0)},\ldots,\theta_m^{(0)})^{T}$, and the optimal calibration parameter $\bm\theta^*=(\theta^*_1,\ldots,\theta^*_m)^T$ by (\ref{thetastar}).
The goal of this paper is to propose a novel calibration method to identify and adjust the sensible variables. {{
		First we consider the estimator defined by the optimization problem}}
{
\begin{multline}
\hat {\bm\theta}=\operatorname*{argmin}_{\bm\theta}(\mathbf Y-\mathbf{Y}^s_{\bm\theta})^{T}( \mathbf \Phi_{\bm\theta}+ \eta^2 I_n)^{-1}(\mathbf Y-\mathbf{Y}^s_{\bm\theta}) + \lambda\sum_{i=1}^m w_i|\theta_i-\theta_i^{(0)}|,\label{eq3.2}
\end{multline}
where $\mathbf Y,\mathbf Y^s_\theta$ and $\mathbf\Phi_{\bm\theta}$ are the same as (\ref{eq2.12}); ${\mathbf W}=\{w_i, i=1,\ldots,m\}$ is a known weights vector; $\lambda$ is a tuning parameter; and $\hat{\theta}=(\hat{\theta}_1,\ldots,\hat{\theta}_m)^T$.}

The objective function of (\ref{eq3.2}) consists of two components.
The first part is 
\begin{eqnarray}
(\mathbf Y-\mathbf{Y}^s_{\bm\theta})^{T}(\mathbf \Phi_{\bm\theta}+\eta^2 I_n)^{-1}(\mathbf Y-\mathbf{Y}^s_{\bm\theta})=:L(\mathbf Y,\mathbf{Y}^s_{\bm\theta}),
\label{eq3.21}
\end{eqnarray}
which is the objective function of the projected kernel calibration method (\ref{eq2.12}).
{{We call this term the \textit{model loss}}}, because it captures the discrepancy between the computer outputs and the physical experimental observations. 
{{The second term is an adaptive-lasso-type penalty. It is a weighted multiple of the $l_1$ distance between the current calibration parameters and the engineering engineering design values, denoted by
\begin{eqnarray}
\begin{aligned}
D_{l_1}(\bm\theta,\bm\theta^{(0)})
=\sum_{i=1}^m w_i |\theta_i-\theta_i^{(0)}|.
\end{aligned}
\label{eq2.13}
\end{eqnarray}
In the statistical analysis, $l_1$-type penalties are widely used because they encourage sparse solutions to break the curse of dimensionality \citep{buhlmann2011statistics,tibshirani1996regression}. Similar with \cite{zou2006adlasso}, we suppose the weight $w_i=1/|\hat {\bm\theta}_i^{or}-\bm\theta_i^{(0)}|$, if $|\hat {\bm\theta}_i^{or}-\bm\theta_i^{(0)}|\neq 0$ and $w_i=\infty$, if $|\hat {\bm\theta}_i^{or}-\bm\theta_i^{(0)}|= 0$, where $\hat {\bm\theta}^{or}=\operatorname*{argmin}_{\bm\theta}L(\mathbf Y,\mathbf{Y}^s_{\bm\theta})$.}}

In the current context, a sparse $\hat{\bm\theta}$ means that $\hat{\theta}_i=\theta^{(0)}_i$ for some or most $i$'s. The degree of sparsity is determined by the tuning parameter $\lambda$. If $\lambda=0$, the proposed method goes back to the projected kernel calibration, which gives a non-sparse solution. When $\lambda$ goes to infinity, the proposed method gives a fully sparse solution with $\hat{\bm\theta}=\bm\theta^{(0)}$. 
In practice, one should make a suitable choice of $\lambda$ to balance the model fitting and the model complexity. 
{{Cross-validation is widely used in identifying the tuning parameter \citep{Trevor09}. However, the adaptive lasso-type variable selection models often includes too many variables when selecting the tuning parameter by cross-validation method. 
\cite{wang2007regression} and \cite{wang2007tuning} demonstrated that adaptive lasso can identify the true model consistently when the tuning parameters is selected by a BIC-type criterion.
As a result, in order to better convey the characteristics of the proposed method, we select $\lambda$ using the BIC-type criterion:
\begin{eqnarray}
\begin{aligned}
BIC_{\lambda}=\log \left(\frac{L(\mathbf Y,\mathbf{Y}^s_{\hat{\bm\theta}_{\lambda}})}{n}\right)+|S_{\lambda}|\times \frac{\log n}{n},
\end{aligned}
\label{bic}
\end{eqnarray}
where $L(\mathbf Y,\mathbf{Y}^s_{\hat{\bm\theta}_{\lambda}})$ is the model loss; $S_{\lambda}=\#\{|\hat{\theta}_{\lambda,j}-\theta_j^{(0)}|\neq 0, j=1,\ldots,m\}$; $\hat{\bm\theta}_{\lambda}$ is the estimate for some chosen value of $\lambda$.
}}

{{As outlined in Section \ref{sec:surrogate}}}, the computer code is expensive to run and thus $\mathbf Y^s_{\bm\theta}$ cannot be regarded as a known function of $\theta$. Therefore, it is infeasible to solve (\ref{eq3.2}) directly. To obtain a computationally efficient estimator, we replace the computer response surface $y^s$ with the surrogate model $\hat{y}^s$ introduced in (\ref{eq3.4}).
According to (\ref{eq2.10})-(\ref{eq2.11}), the projected kernel matrix $\Phi_{\bm\theta}$ in (\ref{eq2.10}) now becomes
\begin{eqnarray}
\Phi_{\hat{\mathbf g}}(\mathbf  x_i,\mathbf x_j)=\Phi(\mathbf x_i,\mathbf x_j)-h_{\hat{\mathbf g}}(\mathbf x_i)^{T}H_{\hat{\mathbf g}}^{-1}h_{\hat{\mathbf g}}(\mathbf x_j),
\label{eq3.5}
\end{eqnarray}
with
\begin{eqnarray}
\begin{aligned}
h_{\hat{\mathbf g}}(\mathbf x)=\int_{\Omega}\hat{\mathbf g}(\mathbf x')\Phi(\mathbf x',\mathbf x)d\mathbf x',\\
H_{\hat{\mathbf g}}=\int_{\Omega}\int_{\Omega}{\hat{\mathbf g}(\mathbf x')}\hat{\mathbf g}(\mathbf x)^{T}\Phi(\mathbf x',\mathbf x)d\mathbf x'd\mathbf x.
\end{aligned}
\label{eq3.6}
\end{eqnarray} 
{{Then the estimator (\ref{eq3.2}) becomes 
\begin{multline}
\hat {\bm\theta} =\operatorname*{argmin}_{\bm\theta} (\mathbf Y-\hat {\mathbf{Y}}^s_{\bm\theta})^{T}(\mathbf \Phi_{\hat{\mathbf g}}+\eta^2 I_n)^{-1}(\mathbf Y-\hat{\mathbf{Y}}^s_{\bm\theta}) +\lambda\sum_{i=1}^m w_i | \theta_i-\theta_i^{(0)}|,
\label{eq3.7}
\end{multline}
where $\hat{\mathbf{Y}}^s_{\bm\theta}=\{\hat y^s(\mathbf{x}_1,\bm\theta),\ldots,\hat y^s(\mathbf{x}_{n},\bm\theta)\}$. We call the estimator defined by (\ref{eq3.7}) the \textit{penalized orthogonal calibration}, and write it as $\hat{\bm\theta}^{PO}$.
}}

{{Because $\mathbf{\Phi}_{\hat{\mathbf g}}$ is a positive semi-definite matrix}, (\ref{eq3.7}) is a convex optimization problem. We can apply the existing methods such as the NEWUOA algorithm \citep{powell2006newuoa} to solve this problem efficiently. In this paper, we use the Monte Carlo method to approximate the integrals in (\ref{eq3.6}).}

{To justify the proposed method,
we examine whether the three types of calibration variables in Table \ref{Table 1} would be adjusted by the method in the desired manner.} 
{{we call the first part of (\ref{eq3.7}), given by}
\begin{eqnarray}
\hat{L}(\mathbf Y, \hat{\mathbf Y}^s_{\bm\theta}):=(\mathbf Y-\hat {\mathbf{Y}}^s_{\bm\theta})^{T}(\mathbf \Phi_{\hat{\mathbf g}}+\eta^2 I_n)^{-1}(\mathbf Y-\hat{\mathbf{Y}}^s_{\bm\theta}),
\label{eq3.8}
\end{eqnarray}
the \textit{empirical model loss}.}{{
First we consider insensitive variables. Suppose $\theta_i$ is insensitive. We also suppose that our surrogate model should be reasonably accurate. Then by definition, $\hat{y}^s_\theta$ should be (nearly) independent of $\theta_i$.
Consequently, $\hat{L}(\mathbf Y,\hat{\mathbf Y}^s_{\bm\theta})$ is also independent of $\theta_i$. In this case, (\ref{eq3.7}) reduces to minimizing $\lambda w_i|\theta_i-\theta_i^{(0)}|+\hat{L}(\mathbf Y, \hat{\mathbf Y}^s_{\bm\theta_{(-i)}})+\lambda D_{l_1}(\bm\theta_{(-i)},\bm\theta_{(-i)}^{(0)})$, which gives the answer $\hat{\theta}_i=\theta^{(0)}_i$.}} Thus the penalized orthogonal calibration leaves $\theta_i$ unadjusted. If $\theta_i$ is sensitive but insensible, according to the consistency of the projected kernel calibration \citep{tuo2017projected}, the minimizer of $\hat{L}(\mathbf Y,\hat {\mathbf Y}^s_{\bm\theta})$ with respect to $\theta_i$ should be close to $\theta^{(0)}_i$.
In this situation, the $l_1$ penalty will regularize $\theta_i$ and set $\hat{\theta}_i=\theta^{(0)}_i$.{{ If $\theta_i$ is sensible, $\theta^{(0)}_i$ does not give a good model fitting. Then for a suitably chosen $\lambda$, the partial derivative of $\hat{L}(\mathbf Y,\hat{\mathbf Y}^s_{\bm\theta})$ at $\theta_i=\theta^{(0)}_i$ has a magnitude greater than $\lambda w_i$, which is the maximum subgradient of $D_{l_1}(\bm\theta,\bm\theta^{(0)})$ with respect to $\theta_i$. As a consequence, $\theta^{(0)}_i$ cannot be a solution to (\ref{eq3.7}) and therefore $\theta_i$ will be adjusted.}}

{{

{{\section{Numerical Study}
		\label{sec:numerical}
		{{
				In this section, we conduct a simulation study to examine the performance of the proposed method.}}
		Suppose the vector of control variables is $\mathbf x=(x_1,x_2,x_3,x_4)^{T}\in [0,1]^4$ and the true process is
		\begin{eqnarray}
		\begin{aligned}
		\zeta(\mathbf x)=&\frac{x_1}{2}\left[\sqrt{1+(x_2+x_3^2)x_4/x_1^2}-1\right]+(x_1+3x_4)\\
		&\times\exp(1+\sin(x_3)).
		\label{eq5.4}
		\end{aligned}
		\end{eqnarray}
		Suppose the computer model is
		\begin{eqnarray}
		\begin{aligned}
		y^s(\mathbf x,\bm\theta)=&\frac{x_1\left(1+\theta_1\sin(x_1)\right)}{2}\sqrt{1+(x_2+x_3^2)x_4/x_1^2}\\
		&+(x_1+\theta_2x_2^2+3\theta_3x_4)\times\exp(\sin(x_3))+\\
		&+\theta_4x_1+\theta_5x_2^2+\theta_6x_3^2+\theta_{10},
		\label{eq5.5}
		\end{aligned}
		\end{eqnarray}
		where $\bm\theta=(\theta_1,\theta_2,\ldots,\theta_{10})^{T}$ is the vector of the calibration parameters. Let $\mathbf X=\{\mathbf x_1, \ldots,\mathbf x_{50}\}$ be the set of the design points, where $\mathbf x_i$'s are generated independently from the uniform distribution on $[0,1]^4$. The physical experimental observations are generated by
		\begin{eqnarray}
		y_i=\zeta(\mathbf x_i)+e_i,
		\label{eq5.6}
		\end{eqnarray}
		where the observation error $e_i$'s are mutually independent and follow $N(0,0.1^2)$.
		
		
		We suppose the engineering design values of $\bm\theta$ are $\bm\theta^{(0)}=(0,0,1,-0.5,0,0,0,0,0,7)^{T}$.
		Let $\hat{\bm\theta}^{PO}=(\hat {\theta}^{PO}_1,\hat{\theta}^{PO}_2,\dots,\hat{\theta}^{PO}_{10})^{T}$ be the penalized orthogonal calibration of $\bm{\theta}$ given by (\ref{eq3.7}). We use a Gaussian function (\ref{eq2.6}) as the kernel function $\Phi$. To determine the hyper-parameter $\phi$ in (\ref{eq2.6}), we first build a Gaussian-process model to approximate the physical observations and estimate $\phi$ by using the maximum likelihood method.  The BIC-type criterion (\ref{bic}) is used to choose the tuning parameter $\lambda$ in the penalized orthogonal estimations of the sensible variables. We repeat the above simulation procedure 100 times to assess the average performance of the proposed method.


		Recall that the optimal value of $\bm\theta$, denoted as $ \bm\theta^{*}=(\theta^{*}_1,\ldots,\theta^{*}_{10})^{T} $, is defined as
		\begin{eqnarray}
		\bm\theta^{*}:
		=\operatorname*{argmin}_{\bm\theta}\int_{\Omega}   ({\zeta(\cdot)-y^s(\cdot,\bm\theta)})^2 d\mathbf{x}.
		\label{eq3.15}
		\end{eqnarray}
		It is worth noting that (\ref{eq3.15}) can only define the sensitive variables, because with respect to the insensitive variables, the above objective function is plat. Using the standard sensitivity analysis method \citep{sobol2001global}, we identify that $\theta_7$, $\theta_8$, $\theta_9$ are insensitive parameters.
		
		We compare the proposed method with the usual projected kernel calibration estimator, denoted by $\hat{\theta}^{PK}$,
		as well as the ordinary least squares estimator $\hat{ \bm\theta}^{OLS}$ \citep{tuo2015efficient} defined as
		\begin{eqnarray}
		\begin{aligned}
		\hat{\bm\theta}^{OLS}:&=\operatorname*{argmin}_{ \bm\theta}(\mathbf Y-\hat {\mathbf{Y}}^s_{\bm\theta})^{T}(\mathbf Y-\hat{\mathbf{Y}}^s_{\bm\theta}).\\
		\end{aligned}
		\label{ols-theta}
		\end{eqnarray}

		We calculate the \textit{integrated error}, defined as
		$$IE(\bm \hat{\theta})=\int_{\Omega}   ({\zeta(\mathbf x)-y^s(\mathbf x,\bm\hat{\theta})})^2 d\mathbf{x},$$
		 to assess the performance of an estimator $\bm\hat{\theta}$. Clearly, $IE$ measures the discrepancy between the true process an the computer model.
		We compute the Monte Carlo average value of $IE(\bm \hat{\theta})$ for each estimator $\hat{\theta}$ mentioned above. We also include the optimal value $\theta^*$ for comparison. 
		The results from the engineering design $\hat{\bm\theta}^{(0)}$, $\hat{\bm\theta}^{OLS}$, the projected kernel calibration estimation $\hat{\bm\theta}^{PK}$ (\ref{eq2.12}), the penalized orthogonal estimation $\hat{\bm\theta}^{PO}$ (\ref {eq3.7}) and the optimal values of the calibration parameters $\bm\theta^{*}$ are summarized in Table \ref{hattheta}. The $IE$ values and the point estimates for $\theta_1$-$\theta_6$ and $\theta_{10}$ are presented.  We skip the estimates for $\theta_7$, $\theta_8$ and $\theta_9$ because they are insensitive parameters and thus their optimal values do not exist.
		
		\begin{table}[htpb]
			\begin{center}
				\caption{
					The performance of $\bm\theta^{(0)}$, $\hat{\bm\theta}^{OLS}$, $\hat{\bm\theta}^{PK}$, $\hat{\bm\theta}^{PO}$ and $\bm\theta^*$.}
				\label{hattheta}
				\begin{tabular}{c|c|c|c|c|c|c|c|c}
					\hline
					$IE$&Estimator&$\theta_1$&$\theta_2$&$\theta_3$&$\theta_4$&$\theta_5$&$\theta_6$&$\theta_{10}$\\
					\hline
					35.177&$\hat{\bm\theta}^{(0)}$ &$0$ & 0& 1 &-0.5&0 &0&7\\
					24.046&$\hat{\bm\theta}^{OLS}$ &$0.001$ & 0.017& 0.027 &-0.060&-0.707 &0.054&7.523\\
					27.712 &$\hat{\bm\theta}^{PK}$ & $-0.014$& $0.020$& $0.200$&$-0.185$&$-0.712$&$0.0275$&$7.388$\\
					22.628&$\hat{\bm\theta}^{PO}$&$0$& $0$& $0.329$&$-0.545$&$0$&$0$&$6.865$\\
					22.581&${\bm\theta}^{*}$&$0$& $-0.001$& $0.300$&$-0.557$&$-0.001$&$-0.002$&$6.849$\\
					\hline
				\end{tabular}
				
			\end{center}	
		\end{table}

		From Table \ref{hattheta}, we can see that $\hat{\theta}_3^{(0)}$, $\hat{\theta}_4^{(0)}$, $\hat{\theta}_{10}^{(0)}$ are relatively far from their optimal values, while other variables are either identical or close to their optimal values. Thus adjusting only $\hat{\theta}_3^{(0)}$, $\hat{\theta}_4^{(0)}$, $\hat{\theta}_{10}^{(0)}$ is ideal, which is exactly what the proposed method does, according to Table \ref{hattheta}.
		This result implies that the proposed method can successfully identify the sensible variables. 
		
		To assess the estimation accuracy regarding the sensible variables, we consider the relative error defined as 
		$$RE_i=|\hat \theta_i -\theta_i^*|/|\theta_i^*|.$$ The results are summarized in Table \ref{T-com}.
	Clearly, the relative errors of the proposed method are much lower than those of the ordinary least squares and the projected kernel methods. This implies the proposed method can adjust the sensible variables effectively. 
	 \begin{table}[htpb]
		\begin{center}
			\caption{
				Relative errors of the estimators.}
			\label{T-com}
			\begin{tabular}{c|c|c|c}
				\hline
				Variable& $RE_{OLS}$& $RE_{PK}$& $RE_{PO}$ \\
				\hline
				$\theta_3$ & 91\% &   33.33\%& 9.667\%\\
				$\theta_4$ & 89.05\%       &   66.79\%&2.154\%\\
				$\theta_{10}$&9.84\%         &  7.87\%&0.234\%\\
				\hline
			\end{tabular}
			
		\end{center}	
	\end{table}
		
		To better understand the performance of the proposed method, we present the curve of the estimators with respect to $\lambda$. Let $\Delta\hat{\theta}_i=|\hat{\theta}^{PO}_i-\theta_i^{(0)}|, i=1,\ldots, 10$. Figure \ref{Figure 1} plots the average curves of each $\Delta\hat{\theta}_i$ over the 100 simulation runs.

		\begin{figure}[htpb]
			\begin{center}
				\includegraphics[scale=0.7]{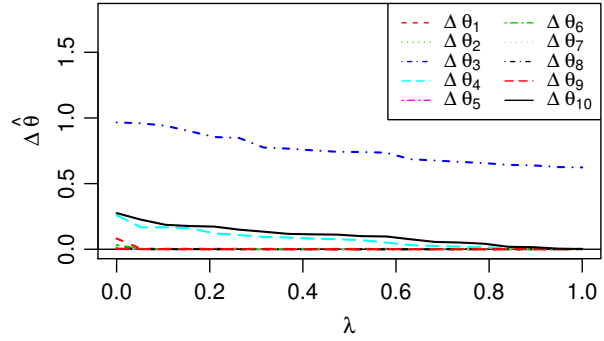}
			\end{center}
			\caption{The relationship between $\lambda$ and $\Delta\hat{\theta}_i$, where $\Delta\hat\theta_i=|\hat \theta_{i}-\theta_i^{(0)}|, i=1,\ldots,10$.  The engineering design value of $\bm\theta$ is $\bm\theta^{(0)}=(0,0,1,-0.5,0,0,0,0,0,7)^{T}$.}
			\label{Figure 1}
		\end{figure}
		
		Figure \ref{Figure 1} shows that,  as $\lambda$  increases, $\Delta\hat\theta_{1}$, $\Delta\hat\theta_{2}$, $\Delta\hat\theta_{5}$, $\Delta\hat\theta_{6}$, $\Delta\hat\theta_{7}$, $\Delta\hat\theta_{8}$ and $\Delta\hat\theta_{9}$ decrease rapidly. Their values vanish even under a small $\lambda$, say $\lambda=0.1$. 
		According to our method, $\theta_{1}$, $\theta_{2}$, $\theta_{5}$, $\theta_{6}$, $\theta_{7}$, $\theta_{8}$ and $\theta_{9}$ are suggested as insensible variables. This is a correct results, as we have learnt from the sensitivity analysis that $\theta_7$-$\theta_9$ are insensitive, and from Table \ref{hattheta} that $\theta_1,\theta_2,\theta_5,\theta_6$ are sensitive but insensible.
		
}}

}}

\section{Composite Fuselage Simulation}
\label{sec:case}

FEA is an effective numerical technique for composite fuselage analysis. During the development of the variation model and quality control system for composite fuselage assembly process, an accurate finite element model is needed, which helps to increase the flexibility and efficiency of model development and control system design. ANSYS Composite PrepPost is an add-in module to the ANSYS Workbench and is integrated with standard analysis for composite parts. By using the Composite PrepPost, two kind of materials, carbon fiber and epoxy resin, are used to generate multiple fabrics with different geometric parameters. Fabrics can be stacked up depending on specific sequence and orientations. And then, stack-ups are used to generate sub-laminates and further integrated into a composite part. Both stress and deviations in addition to a range of failure criteria can be analyzed by using the finite element model.

\begin{figure}[h]
\begin{minipage}[t]{1\textwidth}
\centering
\includegraphics[scale=0.45]{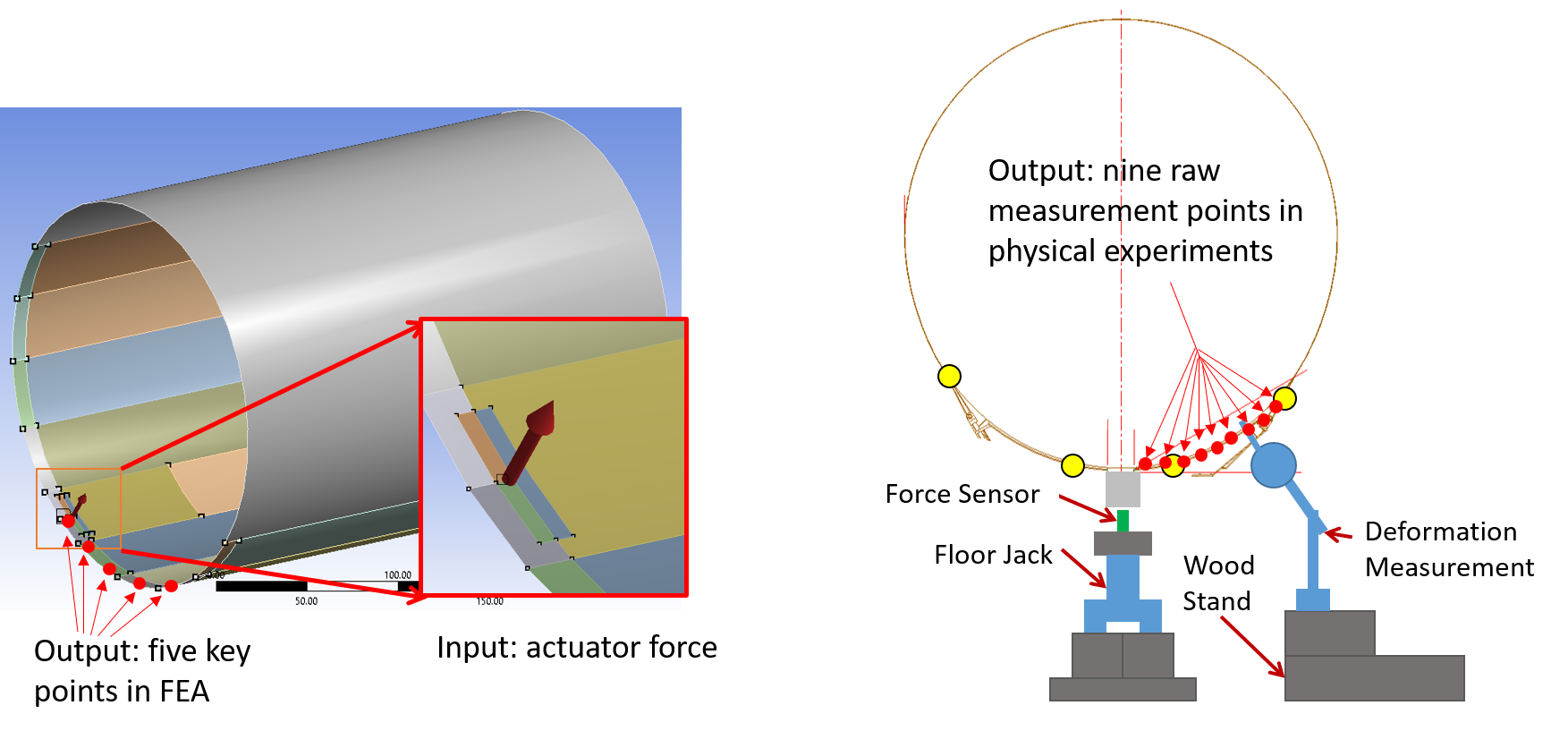}
\end{minipage}%
\caption{FEA simulation: input actuator force and dimensional outputs in five key points (left);  physical experiments setup with dimensional outputs in nine raw measurement points (right).}
 \label{Figure 3}
\end{figure}

In the finite element model of the composite fuselage, input variable, calibration variables, and output variables are summarized in Table \ref{Table 2}. One input variable $x$, actuator force, with eight levels is considered{{, as shown in Figure \ref{Figure 3} left}}. Dimensional deformations in five key points, $ {\mathbf y}^{s}=( { y}^{s}_1,\ldots, { y}^{s}_5)^{T}$, are {{ selected as computer outputs of the FEA}}, shown in Figure \ref{Figure 3} left. The circumferential distances  of these five key points in FEA are $\{0,9.43,18.85,28.27,37.71\}$ inches, respectively. In addition, when we calibrate the finite element model of composite fuselage, five calibration parameters $\theta_1,\ldots,\theta_5$ are considered, including surface body thickness, support parameter, material thickness ratio (between carbon fiber and epoxy resin), fabrics orientation angle, and temperature. {{The engineering design values for calibration parameters $\bm\theta = (\theta_1,\ldots,\theta_5)^{T}$ are determined based on the engineering design of fuselage, or fixture constraints in the physical experiments. However, due to the inevitable manufacturing deviations, the real fuselages cannot be exactly same as design values.}} Here, we obtain the parameters from the literature \citep{feraboli2009characterization} and other engineering background knowledge.

 \begin{table}[h]
\begin{center}
\caption{
Variables in the FEA.}
\label{Table 2}
\begin{tabular}{c|ccccccc}
\hline
Input & actuator force $\mathbf X= \{0,100,200,300,400,500,600,650\}^{T}$\\
\hline
 $\theta_1$      &  surface body thickness, 0.29-0.30 inch \\
   $\theta_2$                & support parameter, 2-4 inches\\
 $\theta_3$                    &  material thickness ratio, 19.5-22.5\\
  $\theta_4$                   &  fabrics orientation, 40-50 degree\\
$\theta_5$&  temperature, 68-71 degree\\
\hline
Output & $ {\mathbf y}^{s}=( { y}^{s}_1,\ldots, { y}^{s}_5)^{T}$\\
\hline
\end{tabular}
\end{center}
\end{table}

 In order to obtain the physical experimental observations, a real structural load experiment is set up to measure the deformations of the composite fuselage under different actuator forces. We ensure that the key parameters (e.g. length, width, thickness, and weight) are consistent between the computer simulated part and the real composite fuselage. In the validation test, the actuator force is changed from 0 to 650 pounds, as shown in Table  \ref{Table 2}. 
 {{Because of the experimental setup, it is challenging to localize the 5 key points of FEA output accurately in our physical experiment. We could obtain the dimensional responses in multiple raw measurement points and then calculate the corresponding observations by linear interpolation. }} In order to cover the range of circumferential distance 0-37.71 inches in the FEA, dimensional deformations in nine points are measured{{, as shown in Figure \ref{Figure 3} right}}.  The circumferential distance of these nine points ranges from about 10 to 45 inches. We need to point out that the circumferential distance ranging from 0 to 10 inches is not measurable because of the fixture set-up of physical experiment. After we get the dimensional deformations in these nine raw measurement points, we compute the five physical experimental observations associated with same circumferential distances in the FEA by linear interpolation. Afterwards, we apply the proposed calibration method based on computer outputs and interpolated physical experimental observations to find the optimal magnitudes of model parameters. 

 All settings details of the finite element model and the physical structural load experiment are shown as follows.

\begin{itemize}
\item Pre-specified design parameter values: $\bm\theta^{(0)}=(0.29,2.5,21,45,69)^{T}$;
\item Physical design: $\mathbf X=\{0,100,200,300,400,500,600,650\}^{T}$;
\item Interpolated physical experimental observations:\\
 $\mathbf {Y}=(\mathbf {Y}_1,\ldots, \mathbf {Y}_5)^{T}; \mathbf {Y}_j=\{y_j(x_1),\ldots, y_j(x_8)\}^{T}, {j=1,\ldots,5}$;
\item Computer design: $\mathbf X^s=\mathbf D\otimes \mathbf X$;
where $\mathbf D=\{\bm\theta_1,\ldots,\bm\theta_{80}\}^{T}$ is a Maximum Latin Hypercube Design \citep{omax1995design} for $\bm\theta$;
\item Computer outputs $\mathbf Y^s_{640\times 5} $.
\end{itemize}

\begin{figure}[htpb]
\begin{center}
\includegraphics[scale=0.7]{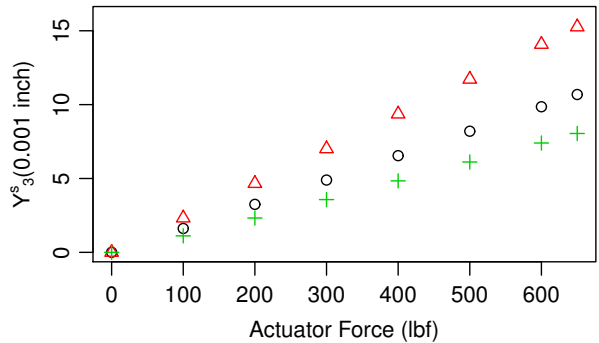}
\end{center}
\caption{Relationship between ${\mathbf X}$ and $\mathbf Y^s_3$ with three different choices of $\bm\theta$: $\bm\theta_1=(0.298,2.3,21.0,45.5,69.4)^T$, $\bm\theta_2=(0.296,2.8,20.4,49.4,70.5)^T$ and $\bm\theta_3=(0.290,2.0,22.1,47.4,68.2)^T$. $\mathbf Y^s_3$ gives the computer outputs in the third observation point.
 \label{Figure 5}}
\end{figure}

Figure \ref{Figure 5} shows the relationship between the computer outputs $\mathbf Y^s_3=(y^s_3(x_1),\ldots, y^s_3(x_8))^{T}$ (associated with circumferential distance 18.85 inches) and the computer design ${\mathbf X}$ with three different choices of $\bm\theta$:

$\bm\theta_1=(0.298,2.3,21.0,45.5,69.4)^T$, $\bm\theta_2=(0.296,2.8,20.4,49.4,70.5)^T$ and $\bm\theta_3=(0.290,2.0,22.1,47.4,68.2)^T$. Figure \ref{Figure 5} shows a linear relationship between $\mathbf X$ and $\mathbf Y^s_3$ for each fixed $\bm\theta$. After a careful examination, we confirm that such a linear relationship between $\mathbf X$ and $\mathbf Y^s_i$ holds for each $i\in\{1,\ldots,5\}$. This result is consistent with our engineering knowledge that the elastic dimensional deformation should be a linear function with respect to the actuator force. Therefore, we use a linear model to link $\mathbf X$ and $\mathbf Y^s_i$'s. The slope of this linear model is a function of $\bm\theta$. Hence, for the input $x$ and $\bm\theta$, we use following model to fit the $j$th item of the computer output:
\begin{eqnarray}
  \hat y_j^s(x,\bm\theta)=h(\bm\theta) x, j=1,\ldots,5.
\label{eq6.1}
\end{eqnarray}
By applying the linear approximation (\ref{eq3.4}), (\ref{eq6.1}) becomes
\begin{eqnarray}
\hat y_j^s(x,\bm\theta)=\hat\beta_{j0} x+\hat\beta_{j1}\theta_1 x+\cdots +\hat\beta_{j5}\theta_5 x, j=1,\ldots,5.
\label{eq6.2}
\end{eqnarray}
The parameters $\beta_{j0},\ldots,\beta_{j5}$ are estimated by using the least squares method.

We calculate the total model loss for all outputs by using a weighted sum of the $j$th empirical model loss $\hat L_j(\mathbf Y,\hat {\mathbf Y}^s_{\bm\theta}), j=1,\ldots,5$ as
\begin{eqnarray}
\begin{aligned}
\hat L(\mathbf Y,\hat {\mathbf Y}^s_{\bm\theta})&=\sum_{j=1}^5 w_j \hat L_j(\mathbf Y,\hat {\mathbf Y}^s_{\bm\theta}),
\end{aligned}
\label{eq6.3}
\end{eqnarray}
where the weight $w_j$ corresponds to the relative importance of this measurement point. In the structural load experiment of the composite fuselage, different measurement points show different quality features, which have weights corresponding to relative importance. According to our engineering knowledge, we choose the weights as
\begin{eqnarray*}
 w_{j}=\exp\{-0.2 (5-j)^2\}, j=1,\ldots,5.
 \end{eqnarray*}
Figure \ref{Figure 6} shows the function relationship between $\lambda$ and $\Delta \hat{\theta}_i, i=1,\ldots, 5$, where $\Delta\hat{\theta}_i:=|\hat{\theta}_i-\theta_i^{(0)}|$:
\begin{figure}[h]
\begin{center}
\includegraphics[scale=0.7]{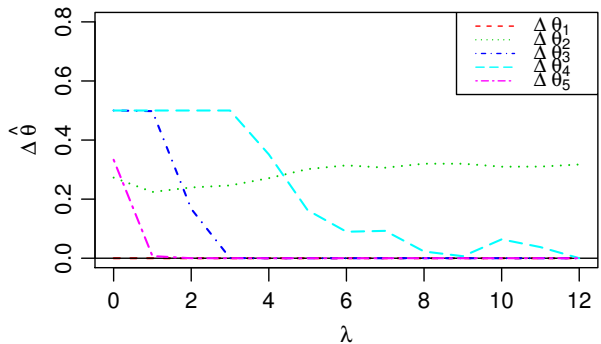}
\end{center}
\caption{The relationship between $\lambda$ and $\Delta\hat{\theta}_i$, where $\Delta\hat\theta_i=|\hat \theta_{i}-\theta_i^{(0)}|, i=1,\ldots,5$. The engineering design value of $\bm\theta$ is $(0.29,2.5,21,45,69)^{T}$.
 \label{Figure 6}}
\end{figure}

It can be seen that $\Delta\hat\theta_1,\Delta\hat\theta_3$ and $\Delta\hat\theta_5$ decay to zero rapidly as $\lambda$ increases, which implies that $\theta_1,\theta_3$ and $\theta_5$ are insensible variables. Consequently, we do not need to adjust the engineering design parameter values of these three parameters. We also observe that $\Delta\hat\theta_2$ does not decay as $\lambda$ increases, and $\Delta\hat\theta_4$ decays rather slowly. Therefore, $\theta_2$ and $\theta_4$ are sensible variables. 
 
\begin{figure}[h]
	\begin{center}
		\includegraphics[scale=0.7]{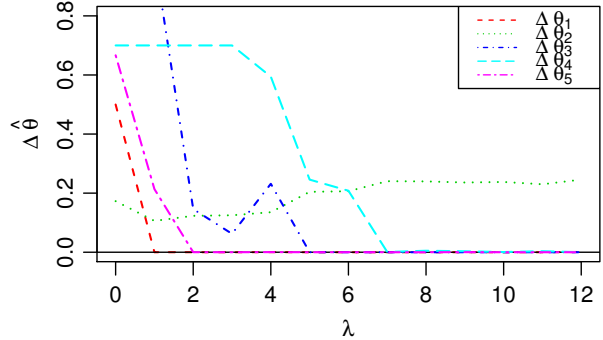}
	\end{center}
	\caption{The relationship between $\lambda$ and $\Delta\hat{\theta}_i$, where $\Delta\hat\theta_i=|\hat \theta_{i}-\theta_i^{(0)}|, i=1,\ldots,5$. The engineering design value of $\bm\theta$ is $(0.295,2.7,22.5,47,70)^{T}$.
		\label{Figure 7}}
\end{figure}

We are also interested in whether $\theta_1,\theta_3$ and $\theta_5$ are sensitive. 
We change the engineering value $\bm\theta^{(0)}$ to $(0.295,2.7,22.5,47,70)^{T}$.
The corresponding results are shown in Figure \ref{Figure 7}.
It can be seen that, in this case,  $\Delta\hat\theta_1$ and $\Delta\hat\theta_5$ decay to zero rapidly as $\lambda$ increases, while $\Delta\hat{\theta}_3$ does not. This implies that, $\theta_1$ and $\theta_5$ are insensitive variables, and $\theta_3$ is sensitive, but its engineering design value $\theta^{(0)}_3=21$ is very close to $\theta^*_3$. 

Next, we will show that, after sensible variable identification, the model loss can be dramatically reduced with a minimum amount of model adjustment. {Table \ref{Table 3} gives the model loss $L(\mathbf Y,\mathbf Y^s_{\bm\theta})$ defined in (\ref{eq3.21}), and empirical model loss $\hat L(\mathbf Y,\hat{\mathbf Y}^s_{\bm\theta})$ defined in (\ref{eq3.8}) under some input settings. It is worth noting that the model loss cannot be computed given only the training data. Here we conduct an extra set of necessary validation computer runs so that $L(\mathbf Y,\mathbf Y^s_{\bm\theta})$ can be computed numerically.} In Table \ref{Table 3}, we list 11 settings of the calibration parameters. The last column of Table \ref{Table 3} gives the number of adjusted calibration parameters, which can be regarded as the model complexity. The first row is the engineering design values $\mathbf{\theta}^{(0)}$. Under this setting, no calibration parameters are adjusted and thus its complexity is zero. Next, we consider the parameter estimation given by the projected kernel method (\ref{eq2.12}) adjusting only sensible variables. Since we have already identified $\theta_2$ and $\theta_4$ as the sensible variables using the proposed method, we consider three cases: (i) the estimation only adjusting $\theta_2$; (ii) the estimation only adjusting $\theta_4$ and (iii) the estimation adjusting both $\theta_2$ and $\theta_4$, denoted as $\hat{\bm\theta}_{(2)},\hat{\bm\theta}_{(4)}$ and $\hat{\bm\theta}_{(2,4)}$ respectively. The model complexities for these three cases are one, one and two, respectively. We also list six calibration parameter values, denoted by ${\bm\theta}_{[i]}  , i=1,\ldots,6$, which are inputs of the training data with the smallest six model losses.


\begin{table}[htpb]
\begin{center}
\caption{
Model loss of different values of $\bm\theta$.}
\label{Table 3}
\begin{threeparttable}
\begin{tabular}{c|c|c|c|c}
\hline
${\bm\theta}$& Values of $\bm\theta$&$L(\mathbf Y,\mathbf Y^s_{\bm\theta})$&$\hat L(\mathbf Y,\hat{\mathbf Y}^s_{\bm\theta})$&Complexity\\
\hline
$\bm\theta^{(0)}$&$(0.290,2.5,21.0,45.0,69.0)^{T}$ &353.15&391.36&0\\
\hline
$\hat{\bm\theta}_{(4)} $&$(0.290,2.5,21.0,40.0,69.0)^{T}$ &212.52&235.31&1\\
$\hat{\bm\theta}_{(2)}$&$(0.290,3.2,21.0,45.0,69.0)^{T}$      & 81.24 &85.41 &1\\
 $\hat{\bm\theta}_{(2,4)}$&$(0.290,3.0,21.0,40.0,69.0)^{T}$      & 53.29 &63.11&2\\
\hline
   $\bm\theta_{[1]}$&$(0.295,2.5,19.6,43.8,68.9)^{T}$                & 52.61&69.38&4\\
 $\bm\theta_{[2]}$&$(0.290, 3.3, 22.5, 42.9, 68.2)^{T}$                & 55.12&73.81&4\\
  $\bm\theta_{[3]}$&$(0.291,2.6	,19.8	,40.0,70.6)^{T}$                   & 57.91&75.57&5\\
$\bm\theta_{[4]}$&$ (0.294,2.4	,20.5	,46.4,70.6	)^{T}$ &65.59&82.81&5\\
$\bm\theta_{[5]}$&$ (0.291,2.4	,19.9	,48.9	,68.3	)^{T}$ &66.50&85.48&5\\
$\bm\theta_{[6]}$&$(0.300,3.4,22.4,47.1,69.7	)^{T}$ &73.50&87.71&5\\
\hline
\end{tabular}
\begin{tablenotes}
\item[1]{$L(\mathbf Y,\mathbf Y^s_{\bm\theta})$ is the model loss (\ref{eq3.21}); $\hat L(\mathbf Y,\hat{\mathbf Y}^s_{\bm\theta})$ is the empirical model loss (\ref{eq3.8}).}
\item[2]{Complexity = the number of parameters been adjusted.}
\end{tablenotes}
\end{threeparttable}
\end{center}
\end{table}

From Table \ref{Table 3}, we can see that before model calibration, the model loss is 353.15. If only $\theta_2$ is adjusted, the model loss decreases to 81.24. When two sensible variables  $\theta_2$ (support parameter) and $\theta_4$ (fabrics orientation) are both adjusted, the model loss further decreases to 53.29. This value is smaller than those given by the most training samples. Only $\bm\theta_{[1]}$ gives a slightly smaller model loss 52.61. However, the complexity for $\bm\theta_{[1]}$ is four, which means one need to adjust two more calibration parameters to achieve a minor improvement in the model loss. Noting the fact that higher model parameter dimensionality results in increasing estimation instability, one should consider $\hat{\theta}_{(2,4)}$ as a better choice of the calibration parameters. { Besides, as we mentioned, the model loss relies on not only the training data, but also the extra validation computer runs. Only the empirical model loss relies solely on the training data. Among all settings shown in Table \ref{Table 3}, the proposed estimator has the smallest empirical model loss. }


\begin{figure}[htpb]
		\includegraphics[scale=0.38]{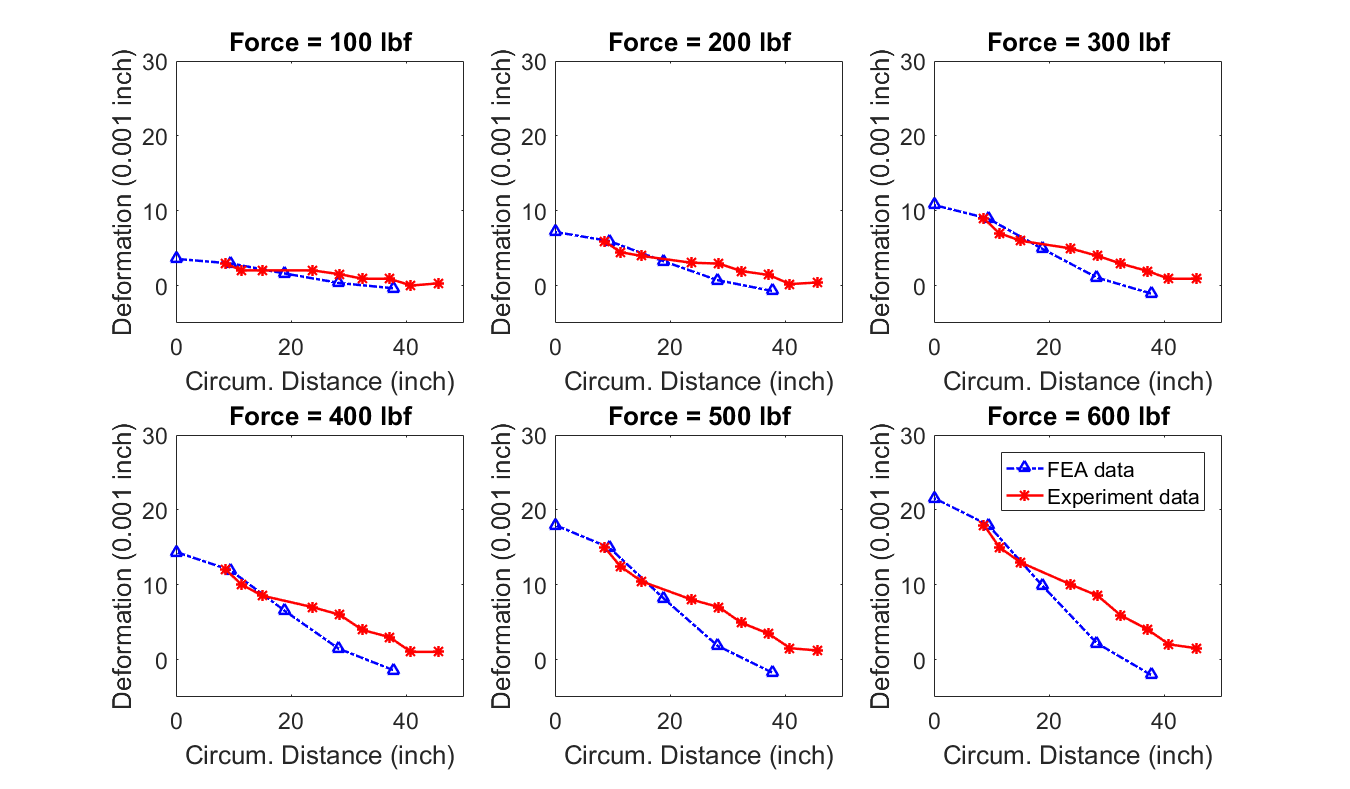}
	\caption{Results before calibration: comparison between computer outputs (FEA data) and physical experimental observations (experiment data) under the actuator force from 100 pounds to 600 pounds.
		\label{Figure 8}}
\end{figure}
\begin{figure}[h]
		\includegraphics[scale=0.38]{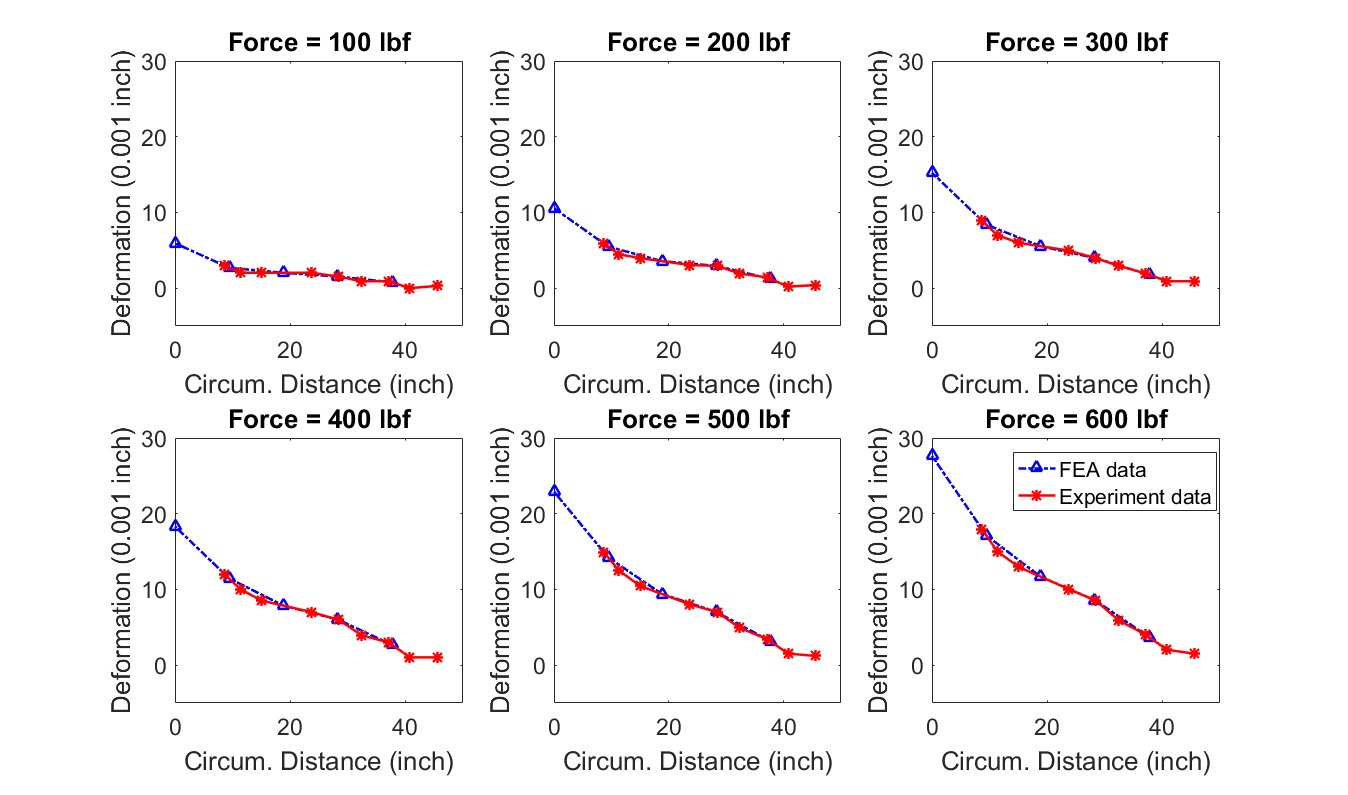}
	\caption{Results after calibration: comparison between computer outputs (FEA data) and physical experimental observations (experiment data) under the actuator force from 100 pounds to 600 pounds.
		\label{Figure 9}}
\end{figure}

We show the effect of model calibration by visual comparison between computer outputs and raw physical experimental observations in Figure \ref{Figure 8} and Figure \ref{Figure 9}. 
Recall that dimensional deformations in nine points are measured in the physical experiment.  So there are nine physical experiment points in Figure \ref{Figure 8} and Figure \ref{Figure 9}.

Figure \ref{Figure 8} corresponds to the $\bm\theta^{(0)}$ with model loss 353.15, while Figure \ref{Figure 9} corresponds to $\hat{\bm\theta}_{(2,4)}$ with model loss 53.29. The horizontal axis is circumferential distance from the center, and the vertical axis is dimensional deformation under the actuator force. From Figure \ref{Figure 8}, we can find that as the circumferential distance becomes larger, the deformation becomes smaller, that makes sense because the actuator force is applied in the center where circumferential distance is zero. However, the discrepancy between computer outputs and physical experimental observations increases when the circumferential distance becomes larger. In addition, as the magnitude of the actuator force becomes larger, the discrepancy between computer outputs and physical experimental observations becomes much larger. In Figure \ref{Figure 9}, the results after calibration are shown. We find that, under different actuator forces, the computer outputs match the physical experimental observations much better.
Although the model fitting is excellent after the calibration, we do not believe that our model overfitts the data because only two calibration parameters are adjusted.

\section{Discussion}
\label{sec:ext}
 Due to inevitable errors or variabilities in manufacturing, the actual values of the physical or engineering attributes of a product can differ from their engineering design values. To build an accurate computer simulator for this product, one needs to calibrate the model parameters.
 A common scenario in computer experiments is that, there exist a number of calibration parameters, while the physical experimental sample size is commonly limited due to certain time and financial constraints. Thus it is often intractable to estimate all calibration parameters using the physical experimental data. 
 
In this paper, we introduce the concept of sensible variables. Adjusting the sensible variables can significantly improve the performance of the computer model. We propose the penalized orthogonal calibration method to identify and adjust the sensible variables. This method can identify the calibration parameters that need to be adjusted. Based on numerical simulation and case study, we show the efficiency and effectiveness of the proposed calibration method. The complex finite element model of the composite fuselage has been improved a lot after calibration, and the computer outputs match the physical experimental observations very well.  

As the computer model can’t match the physical observation perfectly in many cases, the discrepancy between the computer model and the true process can not be ignored. However, in general, there is no bright line between “tuning” and “calibration”. In the proposed method, if the computer model discrepancy is small, then the estimation of model parameters is regarded as calibration instead of model tuning.

\section * {Acknowledgements}

The work is funded by the Strategic University Partnership between the Boeing Company and the Georgia Institute of Technology. Tuo’s work is also supported by NSF grant DMS 1914636.


\bibliographystyle{imsart-nameyear}

\end{document}